\documentclass[reprint,superscriptaddress,amsmath,amssymb,aps,pra,nobibnotes]{revtex4-2}
\usepackage[colorlinks, citecolor = blue, urlcolor = blue]{hyperref}
\usepackage{xcolor}
\usepackage{amsmath}
\usepackage{braket}
\usepackage{cancel}
\usepackage{url}

\usepackage{graphicx}

\usepackage{changes}
\definechangesauthor[color=blue]{AB}
\definechangesauthor[color=red]{MP}

\begin{document}
\definecolor{orcidlogocol}{HTML}{A6CE39}

\title[]{High flux strontium atom source}

\newcommand{\UNIBO}{
Dipartimento di Fisica e Astronomia, Universit{\`a} di Bologna, \\
Via Berti-Pichat 6/2, I-40126 Bologna, Italy}

\newcommand{\LPTWON}{
Université Bordeaux, CNRS, IOGS, LP2N, UMR 5298, F-33400 Talence, France}

\newcommand{\NTU}{
MajuLab, International Joint Research Unit UMI 3654,
CNRS, Universit{\'e} C{\^o}te d'Azur, Sorbonne Universit{\'e},
National University of Singapore, Nanyang Technological University, Singapore}

\newcommand{\NQH}{
Nanyang Quantum Hub, School of Physical and Mathematical Sciences, Nanyang Technological University, 21 Nanyang Link, Singapore 637371, Singapore}

\newcommand{\CQT}{
Centre for Quantum Technologies, National University of Singapore, 117543 Singapore, Singapore}

\newcommand{\Cuno}{\raisebox{.5pt}{\textcircled{\raisebox{-.9pt}{1}}}}
\newcommand{\Cdue}{\raisebox{.5pt}{\textcircled{\raisebox{-.9pt}{2}}}}
\newcommand{\Ctre}{\raisebox{.5pt}{\textcircled{\raisebox{-.9pt}{3}}}}
\newcommand{\Cquattro}{\raisebox{.5pt}{\textcircled{\raisebox{-.9pt}{4}}}}



\author{C.-H. Feng$^{\dagger}$}
\thanks{Now at NPL, London UK}

\author{P. Robert}
\thanks{Contributed equally}
\affiliation{\LPTWON}

\author{P. Bouyer}
\thanks{Now at UvA, Amsterdam NL}
\affiliation{\LPTWON}

\author{B. Canuel}
\affiliation{\LPTWON}

\author{J. Li}
\affiliation{\NTU}
\affiliation{\NQH}

\author{S. Das}
\affiliation{\NTU}
\affiliation{\NQH}

\author{C. C. Kwong}
\affiliation{\NTU}
\affiliation{\NQH}

\author{D. Wilkowski}
\affiliation{\NTU}
\affiliation{\NQH}
\affiliation{\CQT}

\author{M. Prevedelli}
\affiliation{\UNIBO}

\author{A. Bertoldi}
\email{andrea.bertoldi@institutoptique.fr}
\affiliation{\LPTWON}

\begin{abstract}

We present a novel cold strontium atom source designed for quantum sensors. We optimized the deceleration process to capture a large velocity class of atoms emitted from an oven and achieved a compact and low-power setup capable of generating a high atomic flux. Our approach involves velocity-dependent transverse capture of atoms using a two-dimensional magneto-optical trap. To enhance the atomic flux, we employ tailored magnetic fields that minimize radial beam expansion and incorporate a cascaded Zeeman-slowing configuration utilizing two optical frequencies. The performance is comparable to that of conventional Zeeman slower sources, and the scheme is applicable to other atomic species. Our results represent a significant advancement towards the deployment of portable and, possibly, space-based cold atom sensors.

\end{abstract}

\maketitle

\section{Introduction}

The successful implementation of quantum-based atom sensing in various applied and fundamental contexts \cite{Cronin2009,Ludlow2015,Gross2017} requires a good trade-off between a compact, low-power-consumption package and high performance for each subsystem within the experimental setup. Challenges arise at the atomic source stage when dealing with atomic species possessing low vapor pressures, such as alkaline-earth and alkaline-earth-like atoms. The conventional approach involves using an oven to increase the vapor pressure necessary for cold atom experiments \cite{Schioppo2012}. However, this method typically necessitates a bulky Zeeman slower, coupled with a large magnetic field gradient, to accommodate the broad velocity distribution caused by the high-temperature oven \cite{Phillips1982}.

\begin{figure}
     \centering
     \includegraphics[width=0.5\textwidth]{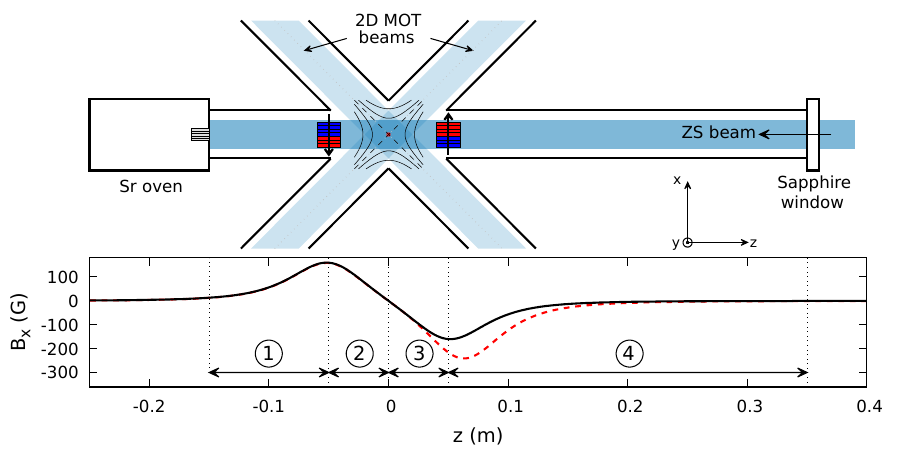}
        \caption{Schematic of the Sr atomic source. (above) An atomic thermal beam is effused by an oven along the $z$ axis; four stacks of permanent magnets (blue and red rectangles) generate a linear quadrupole magnetic field, which vanishes along the $y$ axis and is used to implement a 2D MOT. A laser beam enters the vacuum system along the $z$ axis through a sapphire viewport, and points against the atomic beam. (below) Horizontal magnetic field along the $z$ axis, exhibiting a symmetric magnet configuration with respect to $z=0$ (black, solid line) and when the two stacks of magnets with $z>0$ are displaced from $z=5$ cm to $z=6.23$ cm, and the number of neodymium bars they adopt increased from 9 to 13 (red, dashed line). The four regions used to slow down the atoms are indicated with circled numbers.}
        \label{fig:1}
\end{figure}

A potential solution to achieve miniaturization of the cold atomic jet is to replace the oven with a resistively heated dispenser \cite{Kwon2023} or laser-controlled atomic desorption from a solid atomic source \cite{Kock2016,Hsu2022}. Still, the large velocity dispersion requires to either slow atoms or select a small velocity class of atoms. While a Zeeman slower (ZS) is the most common configuration for slowing a thermal atomic beam, it typically exhibits a large size, high power consumption \cite{Yang2015} or alternatively a complex configuration based on permanent magnets \cite{Reinaudi2012}. Additionally, it inherently generates substantial stray magnetic fields, and its design must mitigate the deleterious effects of a direct line of sight of the oven on the downstream ultracold atoms used in precision measurements. These effects include a high flux of photons resulting from blackbody emission, and a high collision rate with atoms in the thermal beam. Measures such as deflecting the atomic beam produced by the ZS \cite{Yang2015,Cheiney2011} and incorporating in-vacuum optical shutters \cite{Kim2017} are commonly employed to address these issues, but they further increase the complexity of the design.

Recent advancements have aimed to replace conventional ZSs with a more compact alternative, leveraging the cold atomic beam generated by a two-dimensional magneto-optical trap (2D MOT) that is transversely loaded by the oven. Notably, this configuration allows for the manipulation of the cold atoms within an ultra-high-vacuum chamber, shielded from the hot oven surfaces and the atomic beam it produces, and situated along the axis of the 2D MOT where the magnetic field is negligible. This scheme has been successfully implemented for lithium (Li) \cite{Tiecke2009}, sodium (Na) \cite{Lamporesi2013}, and strontium (Sr) \cite{Nosske2017,Barbiero2020,Li2022}. Building upon this basic configuration, further improvements have been achieved by realizing a compact ZS based on the magnetic field slope created by the permanent magnets generating the 2D MOT linear quadrupole \cite{Lamporesi2013,Nosske2017}, implementing sideband-enhanced cooling within the 2D MOT \cite{Barbiero2020}, and utilizing a two-color laser beam to enhance 2D MOT loading efficiency \cite{Li2022}. Despite these solutions, the resulting atomic source only surpasses a standard ZS in terms of compactness and power dissipation, not in atomic flux.

In this work, we present an improved compact atomic source of cold Sr, which produces a state-of-the-art atomic flux, comparable to that of a standard ZS. This result is explained by two main features: Firstly, the distance between the oven and the initial collection region is significantly reduced by up to one order of magnitude, as in \cite{Nosske2017,Barbiero2020,Li2022}, resulting in a 100-fold increase in the useful solid angle for the outgoing atomic thermal beam. Secondly, we achieve optimal deceleration of the oven-generated atomic beam across a velocity range comparable to that of standard Zeeman slowers.

This outcome is facilitated by several factors, including the optimal utilization of the magnetic field within the 2D MOT to enable effective atomic deceleration despite the compact nature of the system. Furthermore, we employ a cascade operation of two laser frequencies within the ZS to increase the range of velocities being captured by the 2D MOT, and modify the magnetic field configuration to shorten the trajectories of captured atoms thereby mitigating losses caused by the radial expansion of the atomic beam.

To evaluate the expected atomic flux, we conduct numerical simulations, which are subsequently confirmed through experimental characterization.

\section{Atomic source}
\label{sec:atomicSource}

Our atomic source, depicted schematically in Figure \ref{fig:1}, bears a close resemblance to the setups described in previous works \cite{Lamporesi2013,Nosske2017,Li2022}. The setup comprises an effusive oven, which generates an atomic thermal beam along the $z$ direction. Four stacks of permanent magnets, each consisting of 9 neodymium bars (Eclipse Magnetics Ltd, model N750-RB, magnetization $M=8.8 \times 10^5$ Am$^{-1}$), are employed to create a linear quadrupole vanishing along the $y$ axis at a distance of 15 cm from the oven, and that enables the operation of a 2D MOT. The magnetic field produced by these magnets has been calculated using the Magpylib Python package; on the propagation axis of the thermal beam ($z$ axis) it is transversal, and directed along the $x$ axis. The magnets are positioned at ±5 cm along the z-axis with opposite magnetization, and have a separation of 7.4 cm along the y-axis.

To slow down the atomic thermal beam, laser light is applied in the $-z$ direction through a sapphire viewport positioned 35 cm away from the point where the magnetic field vanishes, on the side opposite to the oven. The viewport is heated at 380$^{\circ}$C to avoid its metallization with the thermal beam effused by the oven. Additionally, a weak push beam (P$\simeq$100 $\mu$W, resonant with the blue cooling transition) is employed along the $y$ axis to enhance the transfer of atoms, which have undergone pre-cooling in the 2D MOT, to a second cell. This transfer occurs through a vacuum passage 2 mm in diameter, 22.8 mm length, providing a differential vacuum pressure of $\simeq 10^3$. The atoms are subsequently captured in a three-dimensional magneto-optical trap (3D MOT) operating under ultra-high vacuum (UHV) conditions and centered at 35 cm from the output of the differential vacuum stage.

To date, two distinct regions have been considered for implementing a compact ZS \cite{Nosske2017}. The first region, denoted as $\Cuno$ in Figure \ref{fig:1}, encompasses the portion of the setup with a positive magnetic field gradient extending from the oven to the position of the first set of magnets. This region spans the range $-150~\rm{mm}<z<-50~\rm{mm}$ and has a length of $\Delta l_1=100$ mm. The second region, denoted as $\Cdue$, comprises the portion with a negative magnetic field gradient from the first set of magnets to the point where the magnetic field becomes zero. This region covers the range $50~\rm{mm}<z<0~\rm{mm}$ and has a length of $\Delta l_2 = 50$ mm. Since the magnetic field is orthogonal to the propagation direction of the atomic beam, only half of the optical power of the Zeeman cooling beam is effectively utilized for atom deceleration \cite{Ovchinnikov2012,Lamporesi2013,Nosske2017}. Achieving optimal Zeeman slowing with a maximum deceleration $a_{\rm{max}}=v_r/(2\tau)$ across these regions, where $v_r$ represents the recoil velocity of Sr atoms interacting with photons at 461 nm on the $5s^{2} \, ^{1}$S$_0$ -- $5s \, 5p \, ^{1}$P$1$ transition and $\tau$ denotes the lifetime of the corresponding excited state, would enable to bring at rest atoms with velocities at the oven of up to $v_{\rm{max,1}}=434$ m/s in $\Cuno$ and $v_{\rm{max,2}}=306$ m/s in $\Cdue$.

In our study, we achieve an optimal use of the 2D MOT magnetic field for the Zeeman deceleration, with an associated remarkable increase of the captured velocity interval. In detail, we demonstrate the following findings: i) In the region $\Cdue$+$\Ctre$, which encompasses a distance of $\Delta l_{2+3} = 100$ mm, we are able to effectively utilize the entire negative gradient slope of the magnetic field, i.e. also its portion beyond the 2D MOT. This allows us to capture atoms that pass through the 2D MOT once and undergo trajectory inversion near or beyond the second set of magnet stacks (Sec. \ref{ssec:trajectories}). ii) By employing both polarization components of the cooling beam, we can simultaneously decelerate the adjacent low velocity classes of atoms in both regions $\Cuno$ and $\Cdue$+$\Ctre$ (Sec. \ref{ssec:Bmod}). iii) Furthermore, we can increase the upper capture velocity by incorporating a second frequency component in the cooling beam, to decelerate faster atoms between the oven and the first set of magnets so that they are then decelerated by the first frequency component in the region $\Cdue$+$\Ctre$ (Sec. \ref{ssec:fluxEnhancement}). This additional component enables the effective atomic deceleration from the oven to the second set of magnet stacks, covering a distance of 20 cm. For a strontium atom beam at a temperature of 550$^{\circ}$C and considering again the optimal deceleration $a_{\rm{max}}$ over all the distance, this results in an effective capture velocity $v_{\rm{max,1+2+3}} = 613$ m/s, potentially addressing more than 80\% of the atoms emitted from the oven.

\subsection{Numerical model}
\label{ssec:model}

To accurately assess the atomic trajectories and estimate the 2D MOT capture efficiency $\eta_{\mathrm{1D}}$, we employ a numerical model, which takes into account the magnetic field generated by the permanent magnets as well as the radiation pressure exerted by the slowing beam directed against the atomic flux and the 2D MOT beams in close proximity to the $x=0$ position. Our model is based on certain assumptions, consistent with those presented in \cite{Lamporesi2013}: (i) Each atom is treated as a two-level system; (ii) The atomic motion is considered as one-dimensional (1D); (iii) Interatomic collisions are neglected; (iv) We consider classical radiation pressure and neglect the effects of multiple scattering of photons. However, we do account for the small gravitational pull on the atoms. Moreover, to achieve estimated atomic fluxes that are more consistent with experimental measurements, we next consider two factors which reduce the atomic flux captured by the 2D MOT: the radial expansion of the thermal beam, and the radiative decay to levels outside the cooling transition.

To account for the ballistic radial expansion of the atomic beam in the $x-y$ plane and go beyond a purely 1D treatment of the problem, we incorporate the approach described in \cite{Greenland1985} into our model, effectively obtaining a 3D simulation of the atomic dynamics; we neglect instead the radial heating due to spontaneous emission. This allows us to estimate the loss parameter $\xi_{\mathrm{E}}$ associated with each trajectory. When integrated over the velocity interval nominally trapped in the 2D MOT, $\xi_{\mathrm{E}}$ provides the fraction of atoms effectively lost on the surfaces of the vacuum system, denoted $\Xi_{\mathrm{E}}$. The value of $\xi_{\mathrm{E}}$ depends on various factors, including the atomic parameters, the design of the thermal beam collimating devices, and most notably the time-of-flight of each trajectory.
In our experimental setup, to reduce thermal beam divergence, we employ a square array of about 900 Monel 400 micro-sized nozzle tubes mounted at the aperture of the oven. These micro-tubes have an outer(inner) diameter of 0.4(0.2) mm, a length of 9 mm, and they are heated 20$^{\circ}$C above the oven's temperature to avoid clogging their apertures.

Regarding the radiative losses, we consider every atom decaying from the $5s5p \, ^1\textrm{P}_1$ to the $5s4d \, ^1\textrm{D}_2$ during the cooling process (branching ratio $\eta \simeq 1 / \left ( 5 \times 10^4 \right )$) as lost. This is because once it decays back to the ground level, it will be out of resonance with the cooling beam. The loss parameter $\xi_{\mathrm{R}}$ for each trajectory ending trapped in the 2D MOT is obtained by evaluating the probability to quit the cooling transition before reaching for the first time zero velocity. We remark that in this way we slightly underestimated the radiative losses for the trajectories reaching a turning point at $z>0$. An overall radiative loss parameter $\Xi_{\mathrm{R}}$ is computed over the velocity interval being trapped in the 2D MOT.

To obtain the effective capture efficiency $\eta_{\mathrm{E,R}}$ for the 2D MOT, we integrate over the range of captured velocities. We take into account the oven temperature, which is set at 550$^{\circ}$C (corresponding to an approximate root mean square velocity of $v_{\textrm{rms}} \sim 490$ m/s), to weight the relative population according to the initial Maxwell-Boltzmann distribution. Finally, the losses due to radial expansion and radiative decay are incorporated into the model by considering for each velocity class ending in the 2D MOT a reduced capture efficiency equal to $(1-\xi_{\mathrm{E}}) \cdot (1-\xi_{\mathrm{R}})$.

Our primary goal is to optimize the flux of atoms captured by the 2D MOT. To achieve this, we employ a fourth-order Runge-Kutta method to calculate the trajectories of the atoms starting from the oven output at $z=-15$ cm. The calculation continues until one of the following three conditions is met: the atoms hit the sapphire viewport at $z=35$ cm; they return to the oven at $z=-15$ cm; they are captured by the 2D MOT at $z=0$ cm. We identify the capture by the 2D MOT when the trajectory becomes stationary within the region where the 2D MOT cooling beams overlap.

\begin{figure}
     \centering
     \includegraphics[width=0.5\textwidth]{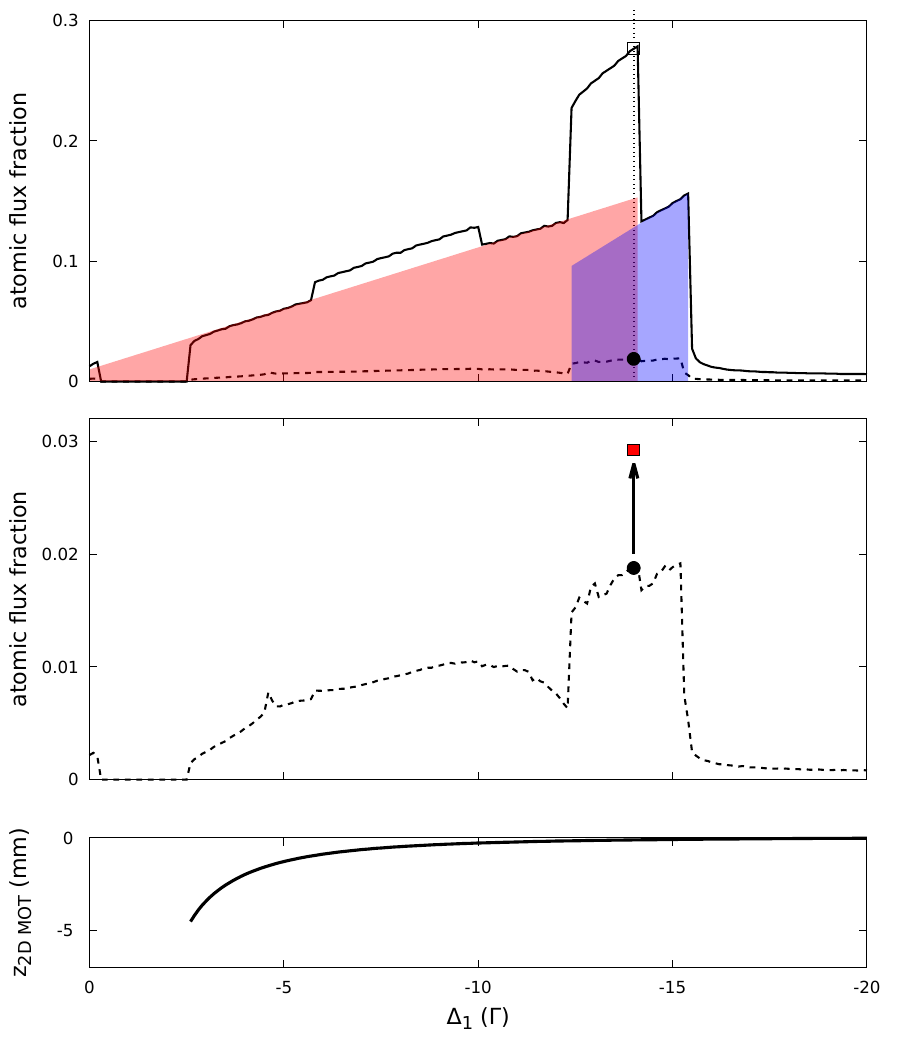}
        \caption{Capture efficiency of the compact ZS. (above) The atomic flux fraction captured by the 2D MOT is calculated versus the detuning $\Delta_1$ of the laser beam used to slow the atoms. The laser intensity is set to 1.5$\, I_{\textrm{sat}}$ on a circular cross section with diameter 9.6 mm: these conditions provide an efficient atomic deceleration on the magnetic field slopes, and comply with the experimentally available laser power. The contribution due to the action of the $\sigma^-$ ($\sigma^+$) polarization of the ZS are highlighted in blue(red). The captured flux is obtained by considering a purely 1D problem (solid line), and by adding the combined losses due to the atomic radial expansion and radiative decay (dashed line); the latter curve alone is depicted more in detail in the central panel. The atomic flux for $\Delta_1 = -14.0 \, \Gamma$ is equal to 27.6\% for the 1D treatment (open black square), reduced to 1.88\% (full black circle) when the contributions of the expansion and radiative losses are taken into account; as indicated by the vertical arrow, the latter value is increased to 2.92\% by the asymmetric magnet configuration (red full square). (below) Displacement of the 2D MOT along the $z$ axis induced by the radiation pressure imposed on the trapped atoms by the slowing beam.}
        \label{fig:2}
\end{figure}

As a reference benchmark, in the absence of any Zeeman slowing beam, approximately 0.41\% of the atoms leaving the oven would be captured by the 2D MOT in the 1D treatment. When accounting for radial losses and considering the 38 mm internal diameter of the vacuum pipe used for the thermal atomic beam, the capture fraction is reduced to $\eta_{\rm{MOT}}=0.11$\%. This reference value gives us an indication of the capture efficiency achievable without any additional measures to slow down the atomic flux. The 2D MOT beams have a $1/e^2$ waist of 12 mm, an optical power of 80 mW, and a detuning of $-1.15 \, \Gamma$ from the cooling transition ($\Gamma=2 \pi \times 30.5$ MHz is the transition natural linewidth), which yields a capture velocity of $\simeq$70 m/s.

The laser beam directed towards the oven along the $z$ direction can be polarized in two main different configurations, taking into account that the magnetic field along the atomic trajectories is primarily directed along $x$. These configurations are as follows: a linear polarization along $x$, which induces $\pi$ transitions in the atoms (e.g., $\ket{^1\textrm{S}_0} \rightarrow \ket{^1\textrm{P}_1, m_F=0}$); a linear polarization along $y$, which induces $\sigma^{\pm}$ transitions in the atoms (e.g., $\ket{^1\textrm{S}_0} \rightarrow \ket{^1\textrm{P}_1, m_F=\pm 1}$). In our setup, we utilize only the second configuration, where the linear polarization is along $y$, inducing $\sigma^{\pm}$ transitions. This arrangement allows for resonance conditions between light and atoms over an extended spatial region due to the varying Zeeman effect associated with a magnetic field slope and the decelerating atomic particles. Efficient slowing of the atoms can be achieved in two ways: they can scatter photons with a $\sigma^-$ polarization along the increasing slope of $B_x(z)$ in region $\Cuno$, or photons with a $\sigma^+$ polarization along the decreasing slope of $B_x(z)$ in region $\Cdue$. Until now, these two conditions have been considered mutually exclusive in previous works \cite{Lamporesi2013,Nosske2017}. As a consequence, in transverse and permanent magnet-based Zeeman slowers (ZSs) \cite{Ovchinnikov2008}, where the magnetic field is orthogonal to the atomic beam, half of the optical power used for atom deceleration is wasted \cite{Ovchinnikov2012}. However, we demonstrate here how to fully utilize the available optical power by addressing different velocity intervals for the atoms effused by the oven with each polarization component.

The slowing beam is linearly polarized along the $y$ axis, so that its power is equally split over the $\sigma^{\pm}$ polarization components relative to the magnetic field. The local acceleration it produces on atoms along the $z$ axis, for a saturation parameter $s$ and red detuning $\Delta_1<0$ with respect to the transition $^1$S$_0$ -- $^1$P$_1$ at 461 nm, is
\begin{align}
  a(z) = - v_r \, \frac{\Gamma}{2} \sum_{m_{\rm{F}}=\pm 1} \frac{s}{1+s+4 \left ( \Delta_{1,\rm{eff}}(z) /\Gamma \right ) ^2},
  \label{eq:a1}
\end{align}
where $\Delta_{1,\rm{eff}}$ depends on $m_{\rm{F}}$, on the atomic velocity $v_{\rm{z}}$ along $z$, and on the local magnetic field $B_{\rm{x}}(z)$:
\begin{equation}
  \Delta_{1,\rm{eff}}(z) = \Delta_1 - k v_z + \frac{\mu_{\rm{B}} \, g_{\rm{F}} \, m_{\rm{F}} \, B_{\rm{x}}(z)}{\hbar},
  \label{eq:deltaEff1}
\end{equation}
where $k$ is the laser wavenumber, $\mu_B$ is the Bohr's magneton, $g_F$ the Landé factor and $h$ the Planck's constant. For the analysis in this section, we defined specific simulation parameters according to the capabilities of our experiment. To ensure effective deceleration on a 40 G/cm slope for atoms with velocities up to 350 m/s, we verified numerically that the saturation parameter for the slowing beam must be at least of 1.5, to ensure a satisfactory scattering rate (the saturation intensity of the $\ket{^1\textrm{S}_0} \rightarrow \ket{^1\textrm{P}_1}$ transition is $I_{\textrm{sat}}=42$ mW/cm$^2$). Increasing $s$ beyond this threshold is inadvisable, because the higher optical power will only generate an increased far-off resonance scattering of photons, with marginal effects on the atomic trajectories. The total optical power available for the deceleration beam is 90 mW, evenly distributed between the two circular polarizations. To verify the condition $s = 1.5$, the cross section for the slowing beam is set to 0.72 cm$^2$. The number of magnets in each of the four stacks is set to 9, which provides a gradient of 40 G/cm at the 2D MOT position. Increasing such value will give access to larger magnetic field slopes, and a potentially more efficient slowing; however, the $s$ required for a larger deceleration will also increase, which will need to find the best trade-off between larger optical power and reduced cross section for the cooling beam.

\subsection{One frequency ZS}
\label{ssec:trajectories}

The atomic flux fraction captured by the 2D MOT is plotted as a function of $\Delta_1$ in the upper panel of Figure \ref{fig:2}. The solid line represents the result of the 1D problem. The dashed line includes the effects of atomic radial spread and radiative decay (represented by parameters $\xi_{\mathrm{E}}$ and $\xi_{\mathrm{R}}$) on each simulated trajectory, providing a more realistic estimate of the captured atoms.

Examining phase-space plots across various $\Delta_1$ confirmed key observations from prior works \cite{Lamporesi2013,Nosske2017,Li2022} and revealed the unique features of our system, as detailed later. The graph in Fig. \ref{fig:2} exhibits two main contributions highlighted with shaded colours: the red one corresponds to the exploitation of the magnetic field slope in regions $\Cdue$ and $\Ctre$ -- and in some cases $\Cquattro$ -- for atomic slowing. It ranges from $\Delta_1=0$ to $\Delta_1=-14.1 \, \Gamma$. In the 1D case, the atomic flux in this region linearly increases from 1.0\% to 15.3\% with a slope of 1.014\%/$\Gamma / 2\pi$, except for three intervals. First, in the range $-2.5 \, \Gamma \le \Delta_1 \le -0.3 \, \Gamma$, the operation of the 2D MOT is destabilized. Second, in the range $-10.0 \, \Gamma \le \Delta_1 \le -5.8 \, \Gamma$, approximately 1.9\% more atoms are slowed down in region $\Cquattro$ by a combination of Zeeman slowing and radiation pressure from the off-resonance beam, enabling their capture by the 2D MOT from the positive $y$ direction. Thirdly, for $\Delta_1 \leq -12.4 \, \Gamma$, the red contribution is combined with the second main contribution, highlighted in blue.
\begin{figure}
     \centering
     \includegraphics[width=0.49\textwidth]{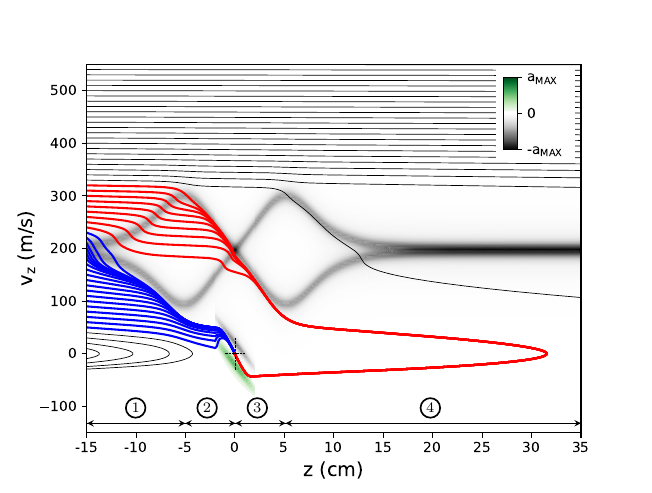}
     \includegraphics[width=0.44\textwidth]{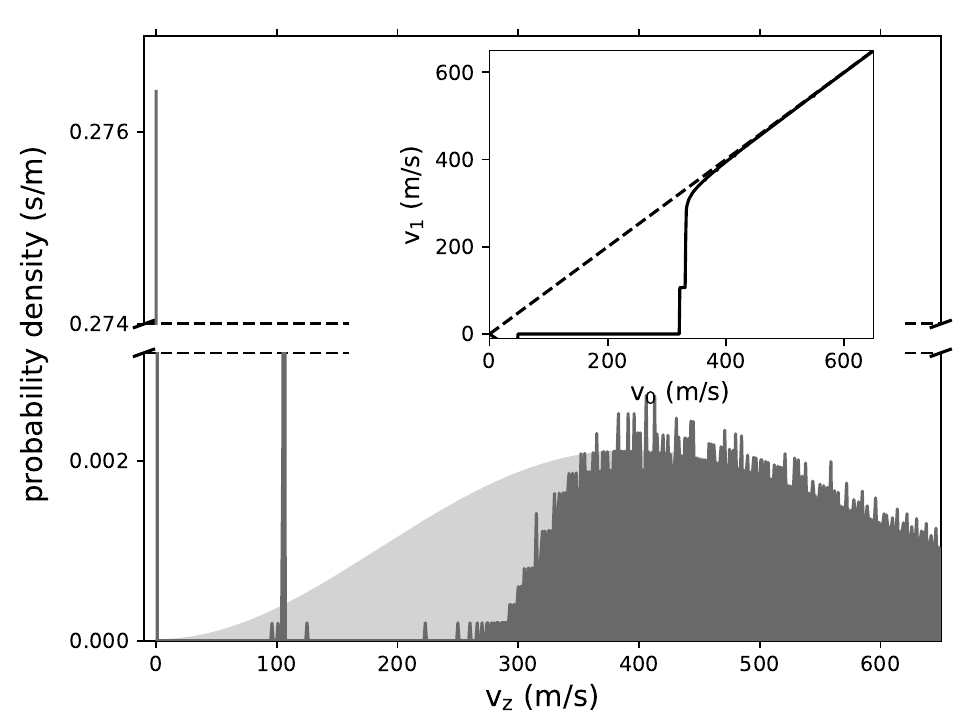}
        \caption{(left) Atomic trajectories from the oven output at $z=-15$ cm, in the presence of a ZS beam with $s=1.5$ and $\Delta_1=-14.0 \, \Gamma$. The acceleration due to the ZS, calculated with Eq. \ref{eq:a1}, and 2D MOT beams is indicated in color code on the phase space diagram. The targeted trapping axis of the 2D MOT at $z$=0 is marked with a dashed cross. The trajectories of atoms captured by the 2D MOT and decelerated by the $\sigma^-$ ($\sigma^+$) polarization of the ZS are traced in blue(red). (right) Probability density versus velocity of atoms emitted by the oven at 550$^{\circ}$C (light gray), and when both the ZS and the 2D MOT are operated (dark gray). The atomic flux captured by the 2D MOT in the 1D case is equal to 27.6\%, indicated by the narrow peak at $v_{\mathrm{z}}=0$ m/s, reduced to 1.88\% when radial and radiative losses are taken into account. Inset: velocity $v_1$ at the end of each trajectory of the ZS as a function of the atomic velocity $v_0$ at the oven output (solid line); the dashed line indicates an unchanged velocity.}
        \label{fig:3}
\end{figure}
The blue contribution extends from $\Delta_1=-12.4 \, \Gamma$ to nearly $\Delta_1=-15.4 \, \Gamma$ and corresponds to the utilization of region $\Cuno$ for decelerating the atoms. In the 1D case, the atomic flux provided by the red contribution linearly increases from 9.6\% to 15.6\% with a slope of 2.00\%/$\Gamma$. This contribution is added to the red contribution for $\Delta_1 \leq 14.1 \, \Gamma$. The slope of the blue area is approximately twice that of the red area: this difference arises because increasing the detuning leads only to a shift upwards of the velocity window captured by the 2D MOT for the red contribution, whereas it also extends the velocity window towards higher velocities for the blue contribution.

The sharp decline in capture efficiency observed at high negative detunings for both color contributions stems from distinct mechanisms. For the red contribution, atomic trajectories end on the viewport where the cooling beam enters. For the blue contribution, slowed down atoms enters the 2D MOT with a velocity too high to be captured.

The lower panel of Fig. \ref{fig:2} shows the vertical displacement of the 2D MOT trapping axis, induced by the radiation pressure exerted by the slowing beam. As anticipated, the displacement approaches zero for larger values of $\Delta_1$. This condition enhances the stability of the experiment since the alignment of the 2D MOT axis must be precise with respect to the small aperture that maintains the differential vacuum between the atomic source chamber and the 3D MOT chamber.

The highest atomic flux is achieved at $\Delta_1=-14.1 \, \Gamma$, corresponding to the high detuning limit, where both polarization components of the Zeeman slower contribute to decelerating the atoms. We investigate the $\Delta_1=-14.0 \, \Gamma$ configuration, which maintains a stability margin of a few MHz, just before an abrupt drop in the 1D atom capture efficiency occurs, as illustrated by the empty square in Fig. \ref{fig:2}. The resulting atomic trajectories are depicted in the upper panel of Fig. \ref{fig:3}, where all four previously defined regions are employed to decelerate specific velocity intervals of the atomic beam emitted by the oven.

In region $\Cuno$, the $\sigma^-$ light decelerates atoms with initial velocities $v_0 \lesssim 230$ m/s to a low speed at $z=-5$ cm. These atoms then gradually decrease their speed due to off-resonant photon scattering, until they are captured by the 2D MOT when approaching from the negative $z$ direction.

In regions $\Cdue$ and $\Ctre$, the $\sigma^+$ light decelerates atoms with initial velocities between 240 m/s and 320 m/s. These atoms pass through the 2D MOT region once without being captured, enter region $\Cquattro$ where they are further slowed down through off-resonant photon scattering from the ZS beam. After being reflected back, they are ultimately captured by the 2D MOT when approaching from the positive $z$ direction.

\begin{figure}
     \centering
     \includegraphics[width=0.49\textwidth]{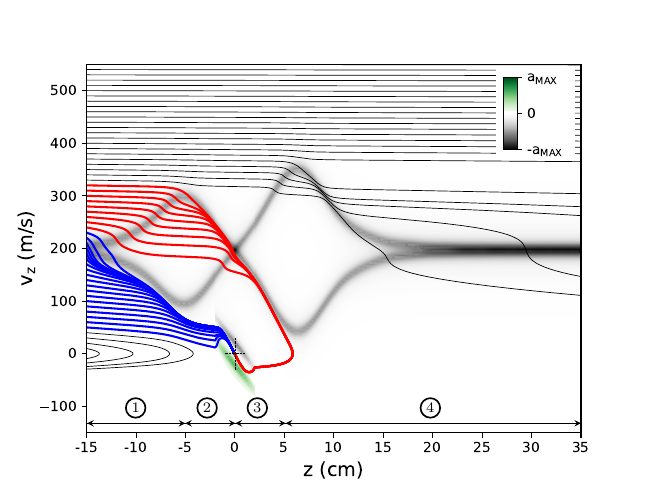}
         \caption{The trajectories extending far on the positive $z$ axis in the configuration of Fig. \ref{fig:3} are brought to an inversion close to the position of the right stacks of magnets, which use 13 instead of 9 elements and are shifted on the $z$ axis so as to maintain the 2D MOT along the $y$ axis. The atomic effective flux captured by the 2D MOT when radial and radiative losses are taken into account raises from 1.88\% of the configuration in Fig. \ref{fig:3} to 2.92\%; this 55\% improvement is depicted in the middle panel of Fig. \ref{fig:2} by the black arrow going from the full black circle to the red full square.}
        \label{fig:4}
\end{figure}

In the right panel of Fig. \ref{fig:3}, the Maxwell-Boltzmann probability density of the atomic beam emitted by the oven at 550$^{\circ}$C is depicted by the light gray shaded area, whereas the dark gray one shows how it is modified by the ZS and the 2D MOT. The inset illustrates the effect on each specific velocity, where the final velocity $v_1$ of each trajectory is plotted against the initial velocity $v_0$ at the oven. The positions considered for determining $v_1$ are $z=z_{\textrm{MOT}}$ for atoms captured by the 2D MOT, $z=-15$ cm for atoms redirected back towards the oven, and $z=35$ cm for atoms reaching the sapphire viewport.

Atoms with velocities between 50 m/s and 320 m/s are captured by the 2D MOT, which corresponds to 27.6\% of the initial distribution (open square in upper panel of Fig. \ref{fig:2}). When considering the radial expansion of the atomic beam, this capture fraction is reduced to 2.79\% (open circle in upper panel of Fig. \ref{fig:2}), and further down to 1.88\% when radiative losses are also taken into account (full circle in upper and middle panels of Fig. \ref{fig:2}). The expansion losses are particularly pronounced for long trajectories resonantly decelerated through regions $\Cdue$ and $\Ctre$. These trajectories spend a long interval at their turning point, located not far from the sapphire viewport at $z=35$ cm, before falling back to the 2D MOT, for a total duration of $\sim$28 ms. For comparison, shorter blue trajectories through regions $\Cuno$ and $\Cdue$ get captured by the 2D MOT in just $\sim$5 ms.

\begin{figure}
     \centering
     \includegraphics[width=0.49\textwidth]{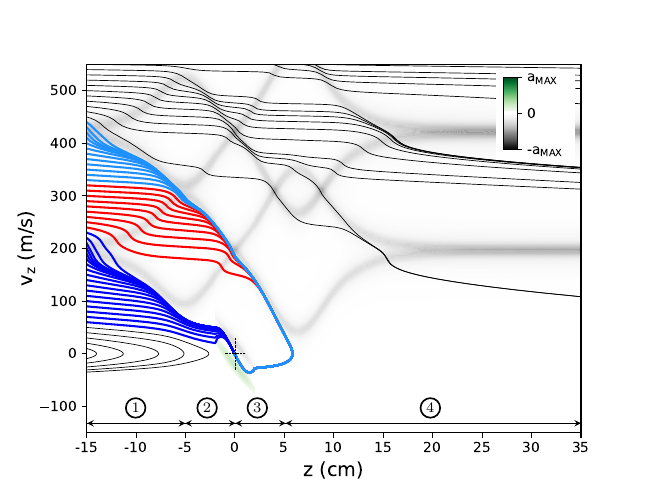}
     \includegraphics[width=0.44\textwidth]{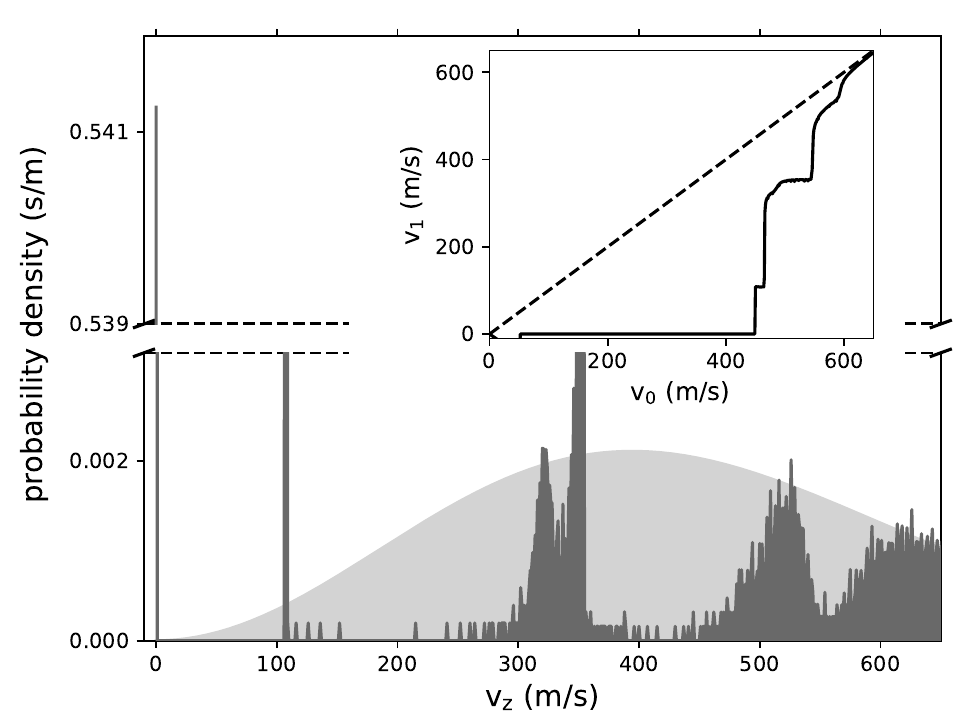}
        \caption{(left) Atomic trajectories from the oven in the presence of a cooling beam with two frequency components, one with a detuning of $-14.0 \, \Gamma$ and $s$=1.5 for each of the two circular polarizations, the other at $=-30.0 \, \Gamma$ and with $s$=2.1. The frequency of the second laser is chosen to decelerate the faster atoms from the oven and up to the second stack of magnets all along, as defined in Eq. \ref{eq:delta2}; the relative trajectories are indicated in light blue. (right) Probability density versus velocity of atoms emitted by the oven at 550$^{\circ}$C (light gray), and after the action of the double frequency ZS and the 2D MOT is taken into account (dark gray). The atomic flux captured by the 2D MOT is equal to 54.12\%, indicated by the narrow peak at $v_{\mathrm{z}}=0$ m/s, reduced to 4.48\% when radial and radiative losses are taken into account. The performance of Fig. \ref{fig:4} is improved by 53\%. Inset: velocity  $v_1$ at the end of each trajectory of the ZS as a function of the atomic velocity $v_0$ at the oven output.}
        \label{fig:5}
\end{figure}

\subsection{Modified magnetic field}
\label{ssec:Bmod}

To shorten long trajectories and minimize radial atom losses, we modify the magnetic field configuration of Fig. \ref{fig:3}. Specifically, the number of magnets in the two stacks at positive $z$ is increased from 9 to 13, and their position is shifted upwards by 1.23 cm by inserting a tailored spacer in the magnets' holder. These adjustments extend the region where atoms are resonantly decelerated on the positive side of the 2D MOT, effectively acting as a sort of ``barrier''. At the same time, the position where the magnetic field is zero, and the magnetic field gradient there remain unchanged, as shown by the dashed red line in the lower panel of Fig. \ref{fig:1}. This modification successfully shortens the previously long trajectories captured by the 2D MOT to a duration of $\sim$7 ms (see Fig. \ref{fig:4}). This leads to a significant 55\% increase in the capture efficiency corrected for radial and radiative losses, raising it from 1.88\% to 2.92\%. Notably, the modified phase space plot justifies a substantial reduction in the length of the vacuum pipe holding the sapphire window, from 35 cm to $\simeq 7$ cm, due to the captured atoms inverting their vertical motion at lower height. In this configuration, the calculated atomic flux reaching and potentially coating the viewport when the 2D MOT is operational increases only by 5\%.

\subsection{Two frequency ZS}
\label{ssec:fluxEnhancement}

Efficiently decelerating higher velocities demands approaches beyond simply scaling up the existing magnet configuration. Increasing the number of magnets proportionally would create larger magnetic field slopes, but the effective deceleration distance will still be roughly 10 cm in both regions. Additionally, limited optical intensity might pose challenges in achieving the required scattering rate for faster atoms to stay resonant with the light throughout the slope. Therefore, exploring alternative deceleration strategies becomes crucial for effectively managing higher velocity atoms.

A promising solution involves using region $\Cuno$ to decelerate atoms with velocities exceeding 320 m/s, bringing them within the effective deceleration range of the ZS light in the regions $\Cdue$ and $\Ctre$. To achieve this, a second optical frequency can be added to the Zeeman cooling beam, with a detuning $\Delta_2$ chosen as
\begin{equation}
  \Delta_2 \simeq \Delta_1 - \frac{2 \, \mu_{\rm{B}} B_{x,\textrm{max}}}{\hbar} - \Gamma \: ,
  \label{eq:delta2}
\end{equation}
where $B_{x,\textrm{max}}$ represents the maximum value of the vertical magnetic field along the $z$ axis near the two stacks of magnets located between the oven and the 2D MOT. The additional term $-\Gamma$ is included to avoid resonantly addressing the same velocities with both optical frequencies simultaneously at $z=-5$ cm. The polarization of this second frequency component is chosen to be linear and aligned along the $y$ axis. This choice ensures a balanced combination of the two circular polarizations, marking a critical distinction from the bi-color scheme implemented in \cite{Li2022}. In the mentioned scheme, the polarization of the second frequency component is oriented linearly along the magnetic field direction, consequently hindering the Zeeman deceleration associated with this beam.

For the second frequency component, we consider a saturation parameter s$_2$=2.1, which guarantees an efficient deceleration on a magnetic field slope of 40 G/cm to atoms with velocities beyond 400 m/s, as we verified numerically. This setup enables the continuous deceleration of the fastest atoms captured by the 2D MOT, from the oven all the way to the position of the second stack of magnets: initially, in region $\Cuno$, they scatter the $\sigma^-$ light component of the second optical frequency; then, in regions $\Cdue$ and $\Ctre$, the $\sigma^+$ light component of the first optical frequency of the ZS beam further decelerates these atoms. Only half the optical power of the second frequency component is exploited for the atom deceleration. As depicted in the left panel of Fig. \ref{fig:5}, this extended deceleration distance spans over more than 20 cm, actually surpassing the separation between the oven and the 2D MOT trapping axis. Consequently, atomic velocities up to 440 m/s can be efficiently trapped, considering the specific experimental configuration under consideration.

Using our numerical tool, we compared the 2D MOT capture efficiency of our cascaded scheme with that of a traditional Zeeman slower using a much longer (50-100 cm) magnetic field slope. The result is a comparable performance, but in a much more compact and power efficient configuration. Furthermore, the near-uniform magnetic field and absence of obstructing magnetic elements in our design make it uniquely well-suited for implementing a radial optical molasses at the oven's output.

\section{Experimental results}

\begin{figure}
     \centering
     \includegraphics[width=0.35\textwidth]{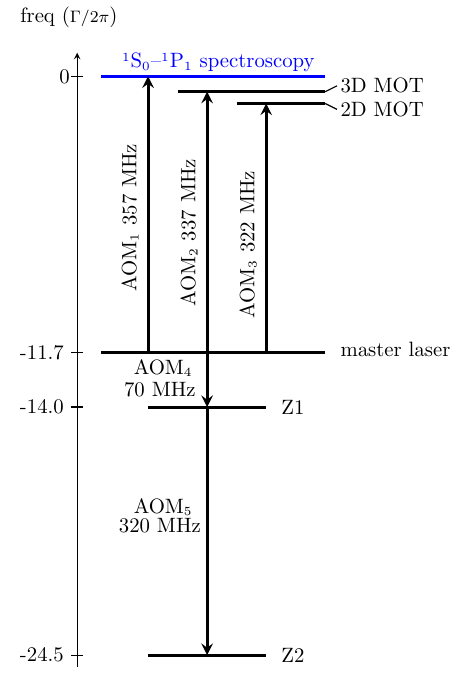}
        \caption{The different laser frequencies needed in the experiment are obtained by using 5 AOMs to shift the master laser light, which is set to $-11.7 \, \Gamma$ from the $^1$S$_0$--$^1$P$_1$ transition used for the spectroscopy (blue horizontal line).}
        \label{fig:6}
\end{figure}

\begin{table*}
\caption{\label{tab:table} Capture efficiency ($\eta_{\mathrm{1D}}$, $\eta_{\mathrm{E,R}}$), losses ($\Xi_{\mathrm{E}}$, $\Xi_{\mathrm{R}}$), and gain with respect to the 2D MOT only ($\eta_{\mathrm{E,R}} / \eta_{\mathrm{MOT}}$) for different experimental configurations of the Sr atomic source, as listed in the three leftmost columns; the capture efficiency when only the 2D MOT is operative is given in the first line of $\eta_{\mathrm{E,R}}$, and is $\eta_{\mathrm{MOT}} \equiv 0.11\%$ . The different lines relates to: (i) No Zeeman cooling beams active, and hence to the performance of the 2D MOT only; (ii) Single Zeeman cooling beam at the optimal detuning, with symmetric magnetic field configuration (see Fig. \ref{fig:3}); (iii) Same as previous, but with asymmetric magnetic field - as in all the following configurations (see Fig. \ref{fig:4}); (iv) two frequencies in the cooling beam, with the second one as implemented experimentally; (v) two frequencies in the cooling beam, both at their optimal value. The uncertainty on the gain factor obtained by the simulation is obtained by considering a $\pm 5\%$ variation on the cooling beam diameter, when the other parameters are maintained fixed; for the gain obtain experimentally the uncertainty results form the fitting procedure on the fluorescence image analysis. All the values refer to the experimental setup considered in the study, and to an atomic source at 550$^{\circ}$C.}
\vspace{0.15cm}
\begin{tabular}{cccc|cccc|cc}
 \hline
 \multicolumn{4}{c}{configuration} & \multicolumn{4}{c}{efficiency and losses} & \multicolumn{2}{c}{$\eta_{\mathrm{E,R}} / \eta_{\mathrm{MOT}}$}\\
 Fig. & $\Delta_1 (\Gamma)$ & $\Delta_2 (\Gamma)$ & \textbf{B} field& $\eta_{\mathrm{1D}} (\%)$ & $\Xi_{\mathrm{E}} (\%)$ & $\Xi_{\mathrm{R}} (\%)$ & $\eta_{\mathrm{E,R}} (\%)$ & simulation & experiment \\ \hline
  \ref{fig:7}a & na & na & sym. & 0.41 & 73 & 0 & $0.11$ & 1 & 1 \\
  \ref{fig:7}b, \ref{fig:3} & $-14$ & na & sym. & 27.6 & 89.9 & 32.6 & 1.88 & 17.1(5) & 14.2(16) \\
  \ref{fig:7}c, \ref{fig:4} & $-14$ & na & asym. & 27.5 & 83.4 & 35.9 & 2.92 & 26.5(7) & 21.3(26) \\
  \ref{fig:7}d & $-14$ & $-24.5$ & asym. & 39.38 & 84.71 & 39.70 & 3.63 & 33.0(8) & 35.4(28) \\
 \ref{fig:5} & $-14$ & $-30.0$ & asym. & 54.12 & 85.16 & 44.20 & 4.48 & 40.7(10) & na \\
\hline
\end{tabular}
\end{table*}

\begin{figure}
     \centering
     \includegraphics[width=0.5\textwidth]{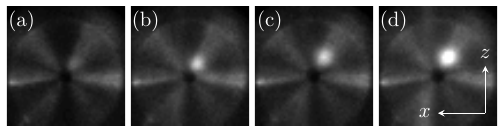}
        \caption{2D MOT observed along the $y$ axis, in four different experimental conditions: no Zeeman slowing beams along the $-z$ direction (a); a cooling beam with frequency set at $-14 \, \Gamma$ from the transition at 461 nm is shined on the atoms along the $-z$ direction, with a symmetric (b; magnetic field profile in Fig. \ref{fig:1}, black solid line) and asymmetric (c; magnetic field profile in Fig. \ref{fig:1}, red dashed line) magnetic field configuration; a cooling beam containing two frequencies at $-14 \, \Gamma$ and $-24.5 \, \Gamma$, respectively, is shined on the atoms along the $-z$ direction, in the asymmetric field configuration (d). The exposure interval for the pictures is 5 ms. The 2D MOT is slightly misaligned with respect to the 2 mm hole used to maintain the differential vacuum between the two chambers and to transfer the cold atoms jet generated by the 2D MOT; this misalignment ensures the visibility of the hole in these pictures.}
        \label{fig:7}
\end{figure}

To experimentally validate the ZS schemes developed with our numerical tool, we implemented a laser system generating all the required optical frequencies. This was achieved using 5 acousto-optic modulators (AOMs) as shown in Fig. \ref{fig:6}. We used a master laser (Toptica, mod. DLC DL pro) locked at $-11.7 \, \Gamma$ from the $^1$S$_0$ -- $^1$P$_1$ cooling transition at 461 nm, and we optically shifted it to generate the cooling and trapping light components. The first Zeeman cooling beam (Z1) was set at the optimal detuning $\Delta_1 = -14.0 \, \Gamma$, whereas the second one (Z2) was shifted to $\Delta_2 = -24.5 \, \Gamma$, \textit{i.e.} which corresponds to a point where the AOM exhibits good diffraction efficiency. To set the frequency at the optimal value of $-30.0 \, \Gamma$ as prescribed by Eq. \ref{eq:delta2}, a different AOM configuration would be required. The 2D MOT beams, and the two beams used for Zeeman cooling were further amplified by optically injecting high-power laser diodes (Nichia, mod. NDB4916). The two Zeeman beams are combined using a 50/50 beam splitter, resulting in a 50\% optical power loss. They are then directed towards the oven along the $z$ axis through the sapphire viewport. The beams are expanded using a common telescope to achieve the following parameters: Z1 an optical power of 45 mW on each polarization, an effective surface of 0.72 cm$^2$ so as to obtain the required intensity of 1.5 times I$_{\mathrm{sat}}$; Z2 an optical power of 75 mW on each polarization, an effective surface of 0.84 cm$^2$ giving an intensity parameter of 2.1. We allocated the laser with the higher output power to the far-detuned frequency component, because it addresses faster atoms and must therefore provide a higher on resonance scattering rate.

To assess the efficiency of the ZS in various configurations, we measured the 2D MOT's fluorescence using a CCD camera oriented along the $y$ direction, which is typically reserved for the push beam used to increase the atomic transfer through the differential vacuum pipe.

Panel (a) of Fig \ref{fig:7} shows the 2D MOT when the ZS is inactive. The subsequent two panels display the signal when the Zeeman cooling beam at $\Delta_1 = -14 \, \Gamma$ is directed at the thermal atom beam effused by the oven, corresponding to the symmetric (b, relative to Fig. \ref{fig:3}) and asymmetric (c, relative to Fig. \ref{fig:4}) magnetic field configurations, respectively. Finally, panel (d) shows the 2D MOT when the second slowing beam Z2, with a detuning $\Delta_2=-24.5 \, \Gamma$ and $s=2.1$, is incorporated into the setup described in Fig. \ref{fig:4}. The brightness of the 2D MOT, when the Zeeman cooling beam is present, allows us to fit the 10 mm$\times$10 mm central region of the related images with the sum of a 2D Gaussian function for the signal originating from the trapped atoms, and a flat offset for the broad fluorescence produced by the atoms while moving along the $z$ axis due to their interaction with the MOT and Zeeman beams. However, this approach was unsuccessful when only the 2D MOT was operating (panel (a) of Fig. \ref{fig:7}), because the signal from trapped atoms was too faint with respect to the background. Instead, we used the picture from panel (b), from which we subtracted the signal from the trapped atoms, to isolate the 2M MOT fluorescence in panel (a). We could thus estimate that the ZS enhances the 2D MOT by a factor of 14.2(16) in the symmetric configuration, 21.3(26) in the asymmetric one, and 35.4(28) in the two cooling beam frequency configuration, as reported in Table \ref{tab:table}. The best enhancement factor we measured is at least a factor 3 better than that reported for similar compact Sr \cite{Nosske2017,Li2022} and Na \cite{Lamporesi2013} sources.

The two cooling frequency configuration involves a cascaded deceleration operation for the high atomic speeds at the oven (light blue trajectories), in combination with deceleration performed on lower initial speeds over 10 cm using only one frequency, either in region $\Cuno$ (blue trajectories) or $\Cdue + \Ctre$ (red trajectories). We confirmed that the dual-frequency ZS system indeed relies on the presence of both ZS beams to function effectively. If Z1 is removed from the dual-frequency configuration, Z2 operates at a frequency where the capture efficiency for a single beam is negligible. The modified setup also retains the advantage of shortening the atomic trajectories through the implementation of a tailored, asymmetric magnetic field. The measured values align reasonably with those obtained by the numerical simulation, as listed in the second-to-last column of Table \ref{tab:table}. Several factors could account for the observed differences: i) Uncertainty in the intensity parameter used in the experiment for the cooling beam, which arises from the relatively low quality of the cooling beam's mode profile; ii) The assumption made in the numerical simulation that radial losses are determined by the expansion from a single microtube, and not from an array, which leads to an overestimation of the effective atomic flux; iii) Heating due to spontaneous emission during the deceleration process.

We also observed a significant reduction in atomic flux at $\Delta_1 = -15 \, \Gamma$, when shifting the working frequency of the AOM controlling the cooling beam. However, this measurement was strongly affected by the AOM operating at the limit of its range. A more precise and comprehensive study of the atomic flux would require a cooling laser capable of being frequency shifted over a wide range without a significant impact on the optical beam's intensity. Such a laser system would offer greater flexibility for exploring various detuning values, optimizing the cooling process, and, most importantly, validating all the results obtained through our numerical analysis.

Boosting the available optical power could significantly increase the cross section of the Zeeman cooling beam, enhancing the 2D MOT's capture efficiency. In a set of experiments, we directed the two frequency components along the $z$ axis without overlapping them on the 50/50 beam splitter, instead imposing a small relative angle estimated to be a few tens of milliradians. However, this approach was unsuccessful, as precise alignment of the two beams proved critical to achieve the cascaded operation of Z1 and Z2.

For a two-frequency ZS with optimal component overlap and minimal optical losses, serrodyne modulation \cite{Johnson2010,Kohlhaas2012} offers a promising approach by generating the second frequency component superimposed with the first one. Given the lack of wideband EOMs in the blue range, the solution could be implemented in a laser system using a fundamental laser at 922 nm, where the saw-tooth modulation would be applied, before doubling in a second harmonic generation (SHG) system to obtain the cooling light at 461 nm.

To finally verify the atomic source efficiency, we put in place the push beam and observed the formation of a 3D MOT in the second vacuum chamber. To this aim, we first precisely aligned the 2D MOT free axis to have it passing through the differential vacuum passage, by acting on precision screws controlling the magnet stacks' $x$-$z$ positions. The trapping region is located at a distance of 35 cm from the differential vacuum pipe, which results in a recapture efficiency of approximately 10\% for the 3D MOT due to the divergence of the pre-cooled atomic beam generated by the 2D MOT. The atomic flux we measured for the 3D MOT is exceeding $5 \times 10^9$ atoms/s for $^{88}$Sr when using the cascade frequency configuration for the ZS; this flux is reduced in accordance with the efficiency gains measured in Table \ref{tab:table} when the atomic source configuration is varied. The best estimated flux at the output of the differential vacuum passage is then comparable to the flux of the commercial cold atomic beam system described in \cite{AOSense}.

\section{Conclusions}
\label{conclusions}

We have implemented several critical improvements in an already compact Sr atomic source, which combines a 2D MOT transversely loaded by a compact ZS. Experimental results are corroborated by a numerical simulation, first developed to reproduce the measured data, and later adopted to evaluate the impact of modified setup configurations and optimize the atomic source efficiency. We make an optimal use of the magnetic field required to operate the MOT, by exploiting both light polarizations to slow down the atoms effused by the oven, by implementing Zeeman slowing all along from the oven to the MOT position and beyond, and by modifying the magnetic field profile to mitigate the issue represented by the thermal radial expansion of the thermal beam produced by the oven. We also operated a novel cascade frequency configuration for the ZS, to extend its capture velocity.

To further increase the atomic flux captured by the 2D MOT one could adopt different solutions. One possibility would be to exploit the almost vanishing magnetic field at the oven's output to implement collimation beams, thus increasing the brightness and collimation of the atomic beam and reducing radial expansion losses. Radiative losses could also be mitigated. Recent research of rapid pumping schemes to efficiently return Sr atoms to the cooling transition at 461 nm predicts that atoms falling in the $^1$D$_2$ level can be pumped back in a very short time to the $^1$S$_0$--$^1$P$_1$ transition using a repumper at 448 nm towards the 5$s \,$8$p \,$$^1$P$_1$ level or at 475 nm towards the 4$d \,$5$p \,$$^1$P$_1$ level \cite{Samland2023}.

The atomic source design, which combines high flux, compactness, and low consumption, could be adapted straightforwardly to other atomic species where a ZS is required, and with a J=0 ground state, such as Ca \cite{Wilpers2002}, Cd \cite{Bandarupally2023}, and Yb \cite{Butler2016,RomeroGonzalez2023}. For atoms with J$\neq$0 ground states like Li, Na, Rb, addressing atomic losses due to optical pumping into non-decelerated magnetic sub-levels remains a challenge, especially near zero magnetic field and, for the two frequency components ZS, where the magnetic field gradient changes sign. The atomic source design could also be integrated in a continuous scheme \cite{Bennetts2017,Cline2022}, contributing to ongoing research in continuous matter wave interferometry \cite{Kwolek2022} and continuous wave superradiant lasing \cite{Kristensen2023}.

\section*{Data availability statement}

All data that support the findings of this study are included within the article.

\section*{Acknowledgments}
\label{sec:acknowledgments}

We thank P. Teulat for the realization of several mechanical components of the experimental setup. P.R. acknowledges funding from the ``Direction Générale de l'Armement" (DGA) for his doctoral fellowship. This work was partly supported by the ``Agence Nationale pour la Recherche" (grant EOSBECMR \# ANR-18-CE91-0003-01, grant ALCALINF \# ANR-16-CE30-0002-01, and grant MIGA \# ANR-11-EQPX-0028), the European Union (EU) (FET-Open project CRYST$^3$, Horizon Europe Qu-Test), IdEx Bordeaux - LAPHIA \# ANR-10-IDEX-03-02 (grant OE-TWC) Horizon 2020 QuantERA ERA-NET (grant TAIOL \# ANR-18-QUAN-00L5-02), Conseil Régional d’Aquitaine (grant USOFF), and Naquidis Center. Qu-Test Project has received funding from the European Union’s Horizon Europe – The EU research and innovation program under the Grant Agreement 101113901.

\section*{References}


\begin{thebibliography}{10}
\expandafter\ifx\csname url\endcsname\relax
  \def\url#1{{\tt #1}}\fi
\expandafter\ifx\csname urlprefix\endcsname\relax\def\urlprefix{URL }\fi
\providecommand{\eprint}[2][]{\url{#2}}

\bibitem{Cronin2009}
Cronin A~D, Schmiedmayer J and Pritchard D~E 2009 {\em Rev. Mod. Phys.\/} {\bf
  81} 1051--1129 \urlprefix\url{https://doi.org/10.1103/revmodphys.81.1051}

\bibitem{Ludlow2015}
Ludlow A~D, Boyd M~M, Ye J, Peik E and Schmidt P 2015 {\em Rev. Mod. Phys.\/}
  {\bf 87} 637--701 \urlprefix\url{https://doi.org/10.1103/revmodphys.87.637}

\bibitem{Gross2017}
Gross C and Bloch I 2017 {\em Science\/} {\bf 357} 995--1001
  \urlprefix\url{https://doi.org/10.1126/science.aal3837}

\bibitem{Schioppo2012}
Schioppo M, Poli N, Prevedelli M, Falke S, Lisdat C, Sterr U and Tino G~M 2012
  {\em Rev. Sci. Instrum.\/} {\bf 83} 103101
  \urlprefix\url{https://doi.org/10.1063/1.4756936}

\bibitem{Phillips1982}
Phillips W~D and Metcalf H 1982 {\em Phys. Rev. Lett.\/} {\bf 48} 596–599
  ISSN 0031-9007 \urlprefix\url{http://dx.doi.org/10.1103/PhysRevLett.48.596}

\bibitem{Kwon2023}
Kwon M, Holman A, Gan Q, Liu C~W, Molinelli M, Stevenson I and Will S 2023 {\em
  Rev. Sci. Instrum.\/} {\bf 94} 013202
  \urlprefix\url{https://doi.org/10.1063/5.0131429}

\bibitem{Kock2016}
Kock O, He W, {\'{S}}wierad D, Smith L, Hughes J, Bongs K and Singh Y 2016 {\em
  Sci. Rep.\/} {\bf 6} 37321 \urlprefix\url{https://doi.org/10.1038/srep37321}

\bibitem{Hsu2022}
Hsu C~C, Larue R, Kwong C~C and Wilkowski D 2022 {\em Sci. Rep.\/} {\bf 12} 868
  \urlprefix\url{https://doi.org/10.1038/s41598-021-04697-4}

\bibitem{Yang2015}
Yang T, Pandey K, Pramod M~S, Leroux F, Kwong C~C, Hajiyev E, Chia Z~Y, Fang B
  and Wilkowski D 2015 {\em Eur. Phys. J. D\/} {\bf 69} 226
  \urlprefix\url{https://doi.org/10.1140/epjd/e2015-60288-y}

\bibitem{Reinaudi2012}
Reinaudi G, Osborn C~B, Bega K and Zelevinsky T 2012 {\em Journal of the
  Optical Society of America B\/} {\bf 29} 729 ISSN 1520-8540
  \urlprefix\url{http://dx.doi.org/10.1364/JOSAB.29.000729}

\bibitem{Cheiney2011}
Cheiney P, Carraz O, Bartoszek-Bober D, Faure S, Vermersch F, Fabre C~M,
  Gattobigio G~L, Lahaye T, Gu{\'{e}}ry-Odelin D and Mathevet R 2011 {\em Rev.
  Sci. Instrum.\/} {\bf 82} 063115
  \urlprefix\url{https://doi.org/10.1063/1.3600897}

\bibitem{Kim2017}
Kim H, Lee W~K, Yu D~H, Heo M~S, Park C~Y, Lee S and Lee Y~K 2017 {\em Rev.
  Sci. Instrum.\/} {\bf 88} 025101
  \urlprefix\url{https://doi.org/10.1063/1.4974756}

\bibitem{Tiecke2009}
Tiecke T~G, Gensemer S~D, Ludewig A and Walraven J~T~M 2009 {\em Phys. Rev.
  A\/} {\bf 80} 013409
  \urlprefix\url{https://doi.org/10.1103/physreva.80.013409}

\bibitem{Lamporesi2013}
Lamporesi G, Donadello S, Serafini S and Ferrari G 2013 {\em Rev. Sci.
  Instrum.\/} {\bf 84} 063102 \urlprefix\url{https://doi.org/10.1063/1.4808375}

\bibitem{Nosske2017}
Nosske I, Couturier L, Hu F, Tan C, Qiao C, Blume J, Jiang Y~H, Chen P and
  Weidem\"{u}ller M 2017 {\em Phys. Rev. A\/} {\bf 96} 053415
  \urlprefix\url{https://doi.org/10.1103/physreva.96.053415}

\bibitem{Barbiero2020}
Barbiero M, Tarallo M~G, Calonico D, Levi F, Lamporesi G and Ferrari G 2020
  {\em Phys. Rev. Appl.\/} {\bf 13} 014013
  \urlprefix\url{https://doi.org/10.1103/physrevapplied.13.014013}

\bibitem{Li2022}
Li J, Lim K, Das S, Zanon-Willette T, Feng C~H, Robert P, Bertoldi A, Bouyer P,
  Kwong C~C, Lan S~Y and Wilkowski D 2022 {\em {AVS} Quantum Science\/} {\bf 4}
  046801 \urlprefix\url{https://doi.org/10.1116/5.0126745}

\bibitem{Ovchinnikov2012}
Ovchinnikov Y~B 2012 {\em Opt. Commun.\/} {\bf 285} 1175--1180
  \urlprefix\url{https://doi.org/10.1016/j.optcom.2011.10.027}

\bibitem{Greenland1985}
Greenland P~T, Lauder M~A and Wort D~J~H 1985 {\em J. Phys. D\/} {\bf 18}
  1223--1232 \urlprefix\url{https://doi.org/10.1088/0022-3727/18/7/009}

\bibitem{Ovchinnikov2008}
Ovchinnikov Y~B 2008 {\em EPJ ST\/} {\bf 163} 95--100
  \urlprefix\url{https://doi.org/10.1140/epjst/e2008-00812-x}

\bibitem{Johnson2010}
Johnson D~M~S, Hogan J~M, w~Chiow S and Kasevich M~A 2010 {\em Opt. Lett.\/}
  {\bf 35} 745 \urlprefix\url{https://doi.org/10.1364/ol.35.000745}

\bibitem{Kohlhaas2012}
Kohlhaas R, Vanderbruggen T, Bernon S, Bertoldi A, Landragin A and Bouyer P
  2012 {\em Opt. Lett.\/} {\bf 37} 1005
  \urlprefix\url{https://doi.org/10.1364/ol.37.001005}

\bibitem{AOSense}
{AOS}ense {C}old {A}tomic {B}eam datasheet
  \url{https://aosense.com/wp-content/uploads/2023/05/AOSense-Cold-Atomic-Beam-System-2023.pdf}

\bibitem{Samland2023}
Samland J, Bennetts S, Chen C~C, Escudero R~G, Schreck F and Pasquiou B 2023
  Optical pumping of 5s4d1d2 strontium atoms for laser cooling and imaging
  \url{https://arxiv.org/abs/2311.02941} (\textit{Preprint}
  \eprint{arXiv:2311.02941})

\bibitem{Wilpers2002}
Wilpers G, Binnewies T, Degenhardt C, Sterr U, Helmcke J and Riehle F 2002 {\em
  Phys. Rev. Lett.\/} {\bf 89} 230801
  \urlprefix\url{https://doi.org/10.1103/physrevlett.89.230801}

\bibitem{Bandarupally2023}
Bandarupally S, Tinsley J~N, Chiarotti M and Poli N 2023 Design and simulation
  of a source of cold cadmium for atom interferometry
  \url{https://arxiv.org/abs/2306.00782} (\textit{Preprint}
  \eprint{arXiv:2306.00782})

\bibitem{Butler2016}
Butler K, Guttridge A, Kemp S, Freytag R, Hinds E~A, Tarbutt M~R and Cornish
  S~L 2016 {\em Rev. Sci. Instrum.\/} {\bf 87} 043109
  \urlprefix\url{https://doi.org/10.1063/1.4945795}

\bibitem{RomeroGonzalez2023}
{Romero González} J, Rahmouni F, Pointard B, Bertoldi A, Lodewyck J and {Le
  Targat} R 2023 {RAZPOUTYNE} project towards compact cooling techniques.
  \url{https://hal.science/hal-04295079/} {FIRST-TF}, Nov 2023, Nice, France.
  (\textit{Preprint} \eprint{ffhal-04295079f})

\bibitem{Bennetts2017}
Bennetts S, Chen C~C, Pasquiou B and Schreck F 2017 {\em Phys. Rev. Lett.\/}
  {\bf 119} 223202
  \urlprefix\url{https://doi.org/10.1103/physrevlett.119.223202}

\bibitem{Cline2022}
Cline J~R~K, Sch{\"a}fer V~M, Niu Z, Young D~J, Yoon T~H and Thompson J~K 2022
  Continuous collective strong coupling between atoms and a high finesse
  optical cavity \url{https://arxiv.org/abs/2211.00158} (\textit{Preprint}
  \eprint{{arXiv:2211.00158}})

\bibitem{Kwolek2022}
Kwolek J~M and Black A~T 2022 {\em Phys. Rev. Appl.\/} {\bf 17} 024061
  \urlprefix\url{https://doi.org/10.1103/physrevapplied.17.024061}

\bibitem{Kristensen2023}
Kristensen S~L, Bohr E, Robinson-Tait J, Zelevinsky T, Thomsen J~W and
  M\"{u}ller J~H 2023 {\em Phys. Rev. Lett.\/} {\bf 130} 223402
  \urlprefix\url{https://doi.org/10.1103/physrevlett.130.223402}

\end{thebibliography}

\providecommand{\newblock}{}

\end{document}